\documentclass[fleqn,twoside]{article}
\usepackage{espcrc2}
 
\usepackage{graphicx}
\usepackage[figuresright]{rotating}

\usepackage{amssymb}
\usepackage{psfig}
\usepackage{times}
\usepackage{natbib}
\usepackage{natbibspacing}
\usepackage{floatflt}

\newcommand{\adspr}{{AdSpR}}
\newcommand{\apj}{{ApJ}}
\newcommand{\apjs}{{ApJS}}
\newcommand{\apjl}{{ApJ}}
\newcommand{\aj}{{AJ}}
\newcommand{\aap}{{A\&A}}
\newcommand{\nat}{{Nat}}

\newcommand{\pasp}{{PASP}}
\newcommand{\npa}{{NuPhA}}
\newcommand{\newar}{{NewA Rev.}}
\newcommand{\iaucirc}{{IAU circ.}} 

\newcommand{\casa}{{Cas A}}
\newcommand{\tiff}{{$^{44}$Ti}}
\newcommand{\scff}{{$^{44}$Sc}}
\newcommand{\caff}{{$^{44}$Ca}}
\newcommand{\cofs}{{$^{56}$Co}}

\newcommand{\arcmin}{{$^{\prime}$}}
\newcommand{\msun}{{$M_{\odot}$}}
\newcommand{\net}{{$n_{\rm e}t$}}
\newcommand{\ep}{{e$^+$e$^-$}}
\newcommand{\fluxunit}{{ph\,cm$^{-2}$s$^{-1}$}}
\newcommand{\kms}{{km\,s$^{-1}$}}

\newcommand{\xmm}{{\it XMM-Newton}}
\newcommand{\chandra}{{\it Chandra}}

\newcommand{\rosat}{{\it ROSAT}}
\newcommand{\einstein}{{\it Einstein}}

\newcommand{\sax}{{\it BeppoSAX}}

\newcommand{\integral}{{\it INTEGRAL}}

\newcommand{\comptel}{{\it CGRO-COMPTEL}}

\title{\centerline{\LARGE X- and $\gamma$-ray Studies of Cas A: 
Exposing Core Collapse to the Core}}
\author{\centerline{\Large Jacco Vink
\address{Columbia University, Columbia Astrophysics Lab., 
550 West 120$^{\rm th}$ St., MC 5247, New York, NY 10027, USA}
\thanks{Chandra fellow}
\thanks{Present address: SRON,Sorbonnelaan 2, 3584 CA Utrecht, The Netherlands}
}}
\begin{document}

\begin{abstract}
In this review of X-ray and gamma-ray observations of
Cas A, evidence is discussed that 
Cas A was a Type Ib supernova of a Wolf-Rayet star 
with a main sequence mass between 22--25~$M_{\odot}$, that exploded
after stellar wind loss had reduced its mass to $\sim6$~$M_{\odot}$.
The observed kinematics and the high $^{44}$Ti yield indicate that
the supernova explosion was probably assymetric, with a kinetic energy of 
$\sim2\times10^{51}$~erg.\\
\vskip 1mm
{\noindent
}
\end{abstract}

\maketitle

\section{Introduction}

\casa\ is the youngest known, and one of the brightest supernova
remnants (SNRs).
It is therefore, arguably, the best galactic SNR to study the
fresh products of explosive nucleosynthesis.

The aim of this review is to discuss the observed properties
in order to put the detection of \tiff\ emission from \casa\
into the more general context of observed nucleosynthesis products, 
inferred explosion energy and progenitor type.
This means I will neglect the equally fascinating topic of cosmic ray 
acceleration by \casa's blastwave \citep[e.g.][]{vink03c}.

\casa, being the brightest radio source in the sky, 
was first discovered in the radio \citep{ryle48}.
Its distance, based on combining Doppler shifts and proper motions,
is $3.4^{+0.3}_{-0.1}$~kpc 
\citep{reed95}, 
at which distance the outer radius of 2.55\arcmin\ corresponds to
2.55~pc. 

The proper motion of optical knots indicate an explosion date around AD 1671 
\citep{thorstensen01}.
This is very close to a putative observation of the supernova by Flamsteed
\citep{ashworth80}. However, \citet{stephenson02} argue that the spurious star
in Flamsteed's catalog, which he observed in AD 1680 and  is 
about 10\arcmin\ away from the position of \casa, 
is not the supernova, but is best explained by assuming that
he mixed up the relative positions of two different stars.

The optical emission consists of fast moving knots, 
characterized by velocities ranging from 
$\sim4000$~\kms\ up to  $\sim15000$~\kms\,
and nitrogen-rich, slow moving knots with typical
velocities of $\sim150$~\kms\, \citep{kamper76}.
The fast moving knots are hydrogen deficient, and dominated by
forbidden O and S emission \citep[e.g.][]{fesen01b}.
The lack of hydrogen-rich ejecta\footnote{There are some exceptions, consisting
of knots with traces of hydrogen and nitrogen, see \citet{fesen91}.} suggests
that \casa\ was a Type Ib SN; the result of the core-collapse
of a Wolf-Rayet (WR) star \citep[see e.g.][]{wlw93}. 

\begin{figure*}
\centerline{
\psfig{figure=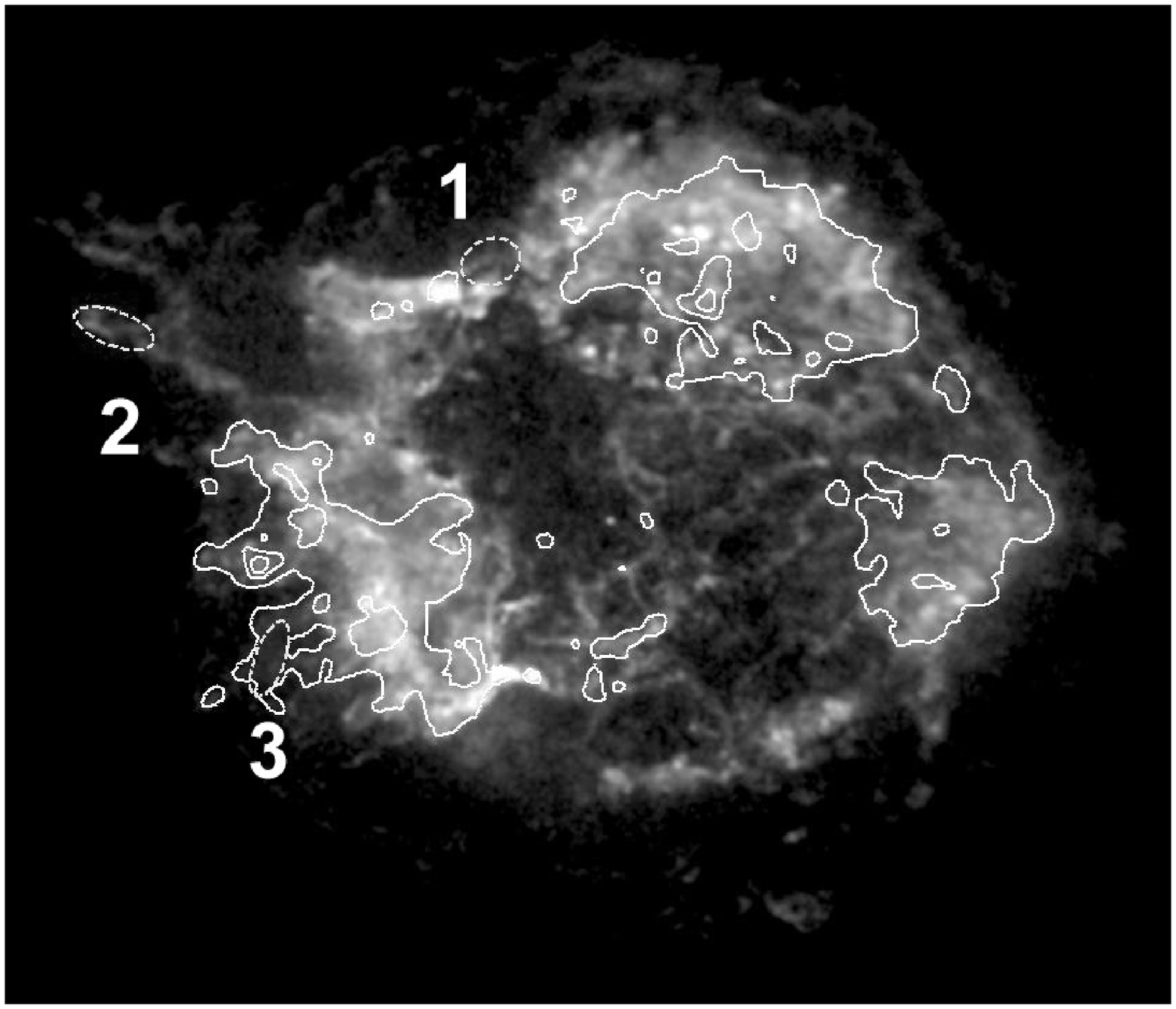,height=6.5cm}
\psfig{figure=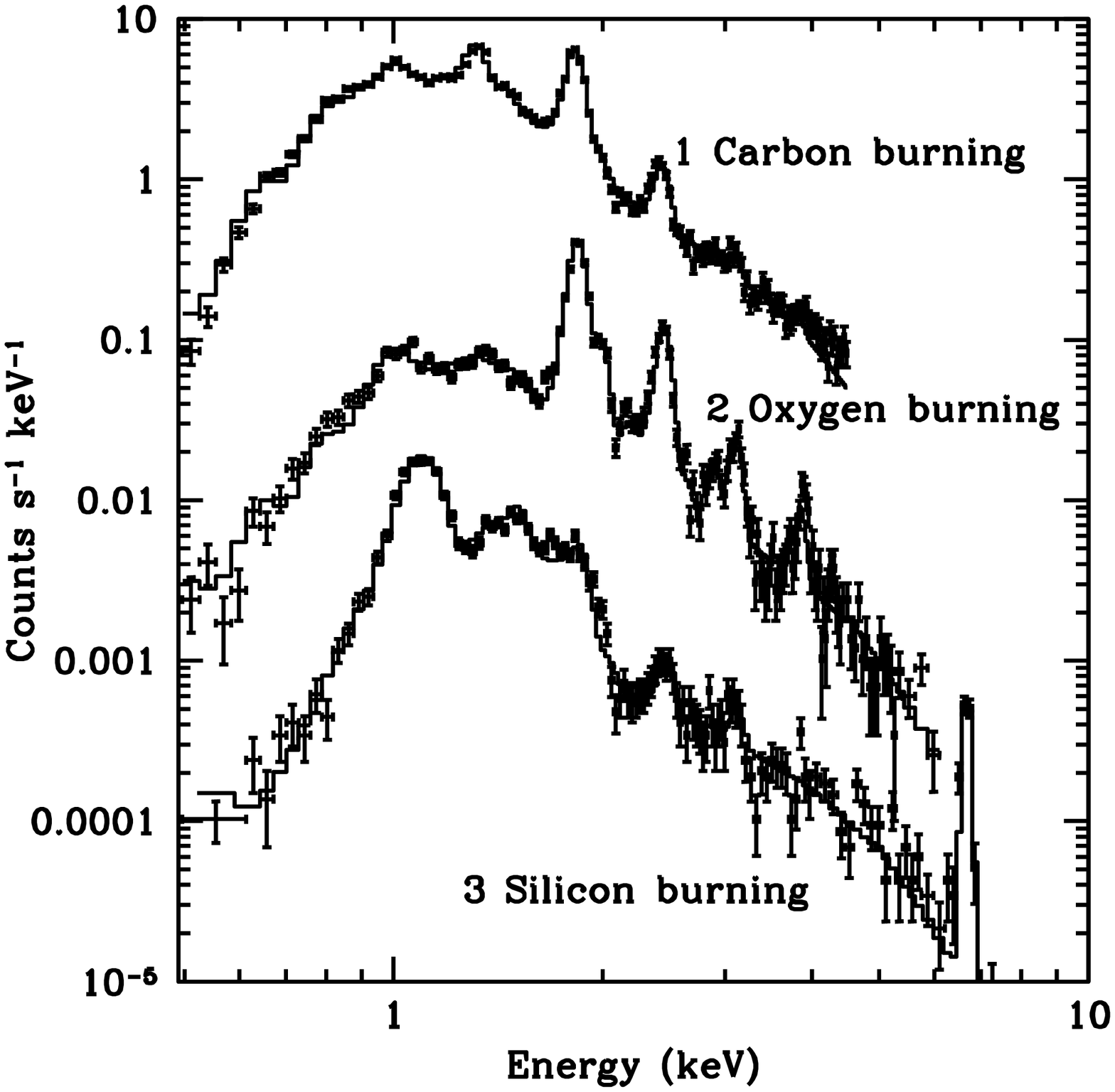,height=6.5cm}
}
\caption{Left: \chandra (ACIS-S3) Si-K image with Fe-K contours overlayed.
The ellipses and labels correspond to 
the extraction regions for the spectra in the right hand figure.
The spectra are labeled according to their dominant nucleosynthesis products.
Note that spectrum no. 3 shows both dominant Fe-L ($\sim$1 keV) and 
Fe-K (6.7~keV) emission.
\label{chandra}}
\end{figure*}

\section{Spatial abundance variations}
X-ray studies of SNRs have the advantage that most of the shocked ejecta
emit X-rays, whereas the optical emitting 
knots constitute only a small fraction of
the ejecta. Moreover, the 0.5-7 keV band covers line emission from
all alpha elements between O and Fe, mostly from H-like and He-like
ions.
This makes it relatively easy to compare the yields of the most
abundant nucleosynthesis products.
This is much harder to do in the optical, where the line
emission is very different from one element to another,
and where some elements (e.g. an abundant element like Si) are 
not observable at all.
Due to the
large absorption toward \casa\ ($\sim 10^{22}$~cm$^{-2}$), 
C and N emission cannot be observed. 

The SNR mass can be estimated from modeling the X-ray spectrum with
a non-equilibrium ionization model. For \casa\ good results
are obtained with a two component model, for which the cooler plasma
component is thought to arise from the shocked ejecta.
The emission measure, together with the assumed or measured abundances,
and an estimate for the volume,
can be directly related to the mass of the plasma components \citep{vink96}.
The lack of hydrogen
in the fast moving optical knots suggests that the X-ray emitting ejecta
are hydrogen deficient as well. This has an important implication for
the ejecta mass, as ionized O, the most abundant element,
is an efficient bremsstrahlung emitter, resulting in a relatively low ejecta
mass estimate, despite the X-ray luminosity of $\sim2\times10^{36}$~erg/s 
\citep{vink96}. 
Taking this into account, current ejecta mass estimates
range from 2-4\msun\, and an O mass of 1--3 \msun\ \citep{vink96,willingale02},
with some uncertainty
due to the uncertain volume filling fraction and temperature structure of 
the ejecta.

The recent advances in X-ray imaging spectroscopy have enabled a detailed
study of spatial abundance variations (Fig.~\ref{chandra}). 
\chandra\ and \xmm\ show that the bright X-ray shell is very rich in
oxygen burning products (Si, S, Ar, and Ca), with a tight correlation
between the abundances of those elements \citep{willingale02}.

Remarkably, although \casa\ is an oxygen-rich remnant, there is a lack
of the carbon-burning products Ne and Mg (Table~\ref{abundances}). 
The Ne/O and Mg/O ratio are almost an order
of magnitude lower than the models of \citet{ww95} (WW95), 
although a few regions show enhanced Ne and Mg emission 
(Fig.~\ref{chandra}).
This is in contrast to other O-rich SNRs like 1E0102.2-72192 
and G292-1.8 \citep[e.g.][]{rasmussen01,park02} that
show prominent Ne and Mg line emission.
The lack of Ne is also apparent from optical observations
\citep{fesen90}.
The reason for this deficiency is not clear; it seems unlikely that
most Ne and Mg is still unshocked, as the ubiquitous Si is formed 
closer to the core of the SN, but, as indicated below, the 
radial ejecta structure may be more complicated than usually assumed.

Even more remarkable is
the fact that in the southeast Fe-rich material exists  outside
the Si-rich shell.
The high Fe abundance clearly indicates that the knots are ejecta 
\citep[Fig.~\ref{chandra}][]{hughes00a,hwang03},
consisting of material from the explosive Si-burning shell, 
where it was synthesized as 
radio-active $^{56}$Ni close to the collapsing
SN core.
The Fe knots must have penetrated the
outer SN layers. In fact, most Fe-K emission is located
outside a radius of 1.45\arcmin, ranging almost out to the blast wave at 
2.55\arcmin, corresponding to average velocities of 4500 - 7800~\kms.
Interestingly the Fe rich knots have an ionization
parameter higher than the Si-rich ejecta
\citep[\net $\sim 5\times10^{11}$\,cm$^{-3}$s,][]{willingale03, hwang03}
indicating that the Fe knots were among the first
to be shocked.\footnote{A plasma is in collisional ionization equilibrium when
\net $> 10^{12}$\,cm$^{-3}$s. For most of the plasma in \casa\ 
\net $\sim 1\times10^{11}$\,cm$^{-3}$ \citep{willingale02}.}
Recently \citet{hwang03} reported the discovery of a 
Fe-rich knot that seems almost devoid of Si. This suggests that at least
this knot is a product of $\alpha$-rich freeze out, 
which results in a lower Si to Fe ratio than explosive Si-burning 
\citep{arnett96}.
This process is also the main source of radio-active
\tiff.
Interestingly, recent 2D simulations of core-collapse supernovae
show that high velocity $^{56}$Ni-rich clumps are formed,
which are not slowed down in Type Ib explosions \citep{kifonidis03}.

\begin{table*}
\begin{center}
\parskip 0mm
\caption{Elemental mass ratios as determined by \citet[][VKB96]{vink96}
and \citet[][W02]{willingale02}, with comparisons to
models by WW95 and WLW93 by \citet{wlw93}.\label{abundances}}
{
\def\arraystretch{1.1}
\begin{tabular}{llllllll}
\noalign{\smallskip}\hline\noalign{\smallskip}
         & VKB96 & W02 & S12A & S22A & S30A & S30B &WLW93\\
\noalign{\smallskip}
\hline
\noalign{\smallskip}
Ne/O & 0.022 & 0.026 & 0.140 & 0.046 & 0.119 & 0.100 & 0.109\\
Mg/O & 0.009 & 0.006 & 0.053 & 0.026 & 0.074 & 0.071 & 0.025\\
Si/O & 0.039 & 0.043 & 0.441 & 0.161 & 0.039 & 0.079 & 0.123\\
S/O  & 0.026 & 0.026 & 0.378 & 0.080 & 0.004 & 0.022 & 0.071\\
Ar/O & 0.008 & 0.007 & 0.127 & 0.015 & 0.001 & 0.004 & 0.014\\
Ca/O & 0.009 & 0.005 & 0.071 & 0.007 & 0.004 & 0.003 & 0.010\\
Fe/O & 0.014 & 0.022 & 0.071 & 0.017 & 0.008 & 0.009 & 0.174\\
\noalign{\smallskip}
\hline
\end{tabular}
}
\end{center}
\end{table*}

\section{The detection of $^{44}$Ti}
Alpha-rich freeze out occurs when, in the expanding Si-burning plasma,
the density drops below the threshold for the triple-$\alpha$ reaction, 
and an excess of $\alpha$-particles results in a relatively large build up of 
\tiff.
Although, its yield is low compared to the 
$^{56}$Ni, \tiff\ is an excellent explosion
diagnostic, as it depends sensitively on the explosion energy and asymmetry, 
and the mass cut \citep{diehl98}.

\tiff\ decays into \scff\ by electron capture with a decay time of 86~yr,
which decays in 
5.7~hr by beta decay into \caff\ \citep[][for the latest measurements]{hashimoto01}.
The decay is therefore long enough
to hope to detect the associated $\gamma$-ray line 
emission from young SNRs at 68~keV and 78~keV
(from excited \scff) and 1157~keV (\caff).
The detection of 1157~keV from \casa\ \citep{iyudin94} 
came nevertheless as a surprise, as the flux observed by \comptel,
$(7\pm1.7)\times10^{-5}$~\fluxunit, implied a 
much higher yield than predicted (WW95).
However, additional observations, and finally a detection of
the \scff\ lines by \sax, indicate a lower flux of 
$(3\pm0.6)\times10^{-5}$~\fluxunit, 
implying a yield of $1.8\times10^{-4}$~\msun\
\citep{vink01a,vink03a}.
This is still on the high side of the predicted yields, 
and is more consistent with models for supernovae with
high explosion energies or asymmetries \citep[WW95,]{nagataki98}. 
The $^{44}$Ti yield for \casa\ may also have been lower,
if, in the course of its lifetime, a substantial
fraction of \tiff\ was ionized beyond the Li-like state 
\citep{motizuki99}.
However, this effect depends sensitively on the ionization history and, 
even if substantial,
is still more in agreement with a higher than expected yield 
\citep{motizuki01,laming01c}.

\begin{figure}
\centerline{
\psfig{figure=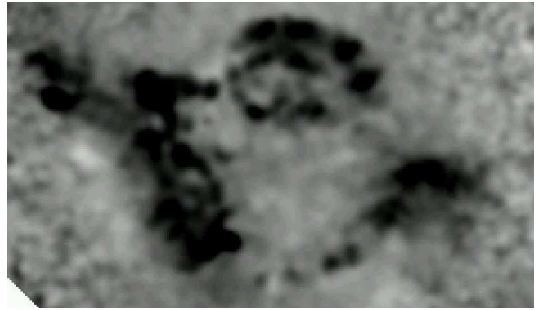,width=7cm}
}
\caption{Ratio map of \chandra\ Si-K and Mg-K images, suggesting
the presence of a counter jet.
\label{jet}}
\end{figure}

\section{\casa's dynamics}
Asymmetric SN Type Ibc explosions have recently
received a lot of attention, as they may be related to
the highly collimated explosions of gamma-ray bursts.
\casa's high \tiff\ yield may be the result of either an
asymmetric or a more energetic explosion. 
Observations support both possibilities. 
The optical morphology and Doppler shifts clearly
show an asymmetric expansion \citep{reed95,lawrence95}, 
but the optically knots are dynamically not very significant,
as they comprise only a small fraction of the total mass 
\citep[$<0.1$~\msun][]{willingale03}.
Most of the shocked plasma is hot enough to be visible in X-rays.
The X-ray emitting material is, therefore, from a dynamical
point of view more important.

The first hints for an asymmetrical expansion of the hot plasma
came from \einstein\ FPCS 
observations of Si and S lines \citep{markert83}, 
showing that the southeast has an overall redshift, 
whereas the northern part is blueshifted. The most comprehensive
X-ray Doppler study is based on \xmm\ data, 
showing that the north/southeast asymmetry is even more apparent
for Fe-K emission, with Doppler velocities up to 2500~\kms\
in the North \citep{willingale02}.
So, interestingly, Fe rich material
is ahead of the Si rich material in ``Doppler space'' in the North, 
whereas in projection Fe is ahead of Si-rich material in the Southeast.
This also indicates that there is no obvious symmetry to the
Fe kinematics; 
there is no evidence for a Fe-rich jet, nor a well defined
Fe-rich equatorial structure.
The jet-like structure in the East is Si-rich (Fig.~\ref{chandra}). 
There is no clear evidence for a counter jet, but mapping the
ratio of Si-K and Mg-K emission does reveal a structure that
hints at the presence of counter jet (Fig.~\ref{jet}).
In fact the \chandra\ images
show Si-rich knots outside the main shell
that seem to have pierced through the denser material in the West.

The rim of continuum emission surrounding \casa\ is likely due
to synchrotron emission associated with the blastwave 
\citep{gotthelf01a,vink03a}. Apart from near the jet region and the West, 
this rim is remarkably circular, much more so than the Si-emitting material.
This is a clear hint that the observed asymmetries are predominantly
due to an asymmetric explosion, as opposed to pre-existing structures in
the circumstellar medium (CSM).

Note that the Doppler shifts are based on
centroid fitting and are possibly affected by line of sight effects,
so that actual velocities may be higher. Based on X-ray proper motions
using \einstein\ and \rosat\
\citet{vink98a} inferred a blastwave velocity for \casa\ of 5200~\kms,
a value that has recently been confirmed with \chandra\ \citep{delaney03}.
Based on the Doppler velocities \citet{willingale03} estimated the explosion
energy of \casa\ to be $10^{51}$~erg. However, they ignored the
high blastwave velocity. If that is taken into account, an explosion
energy of $2\times10^{51}$~erg can be inferred. Hydrodynamic modeling
of the blast wave velocity and the blastwave and reverse
shock radii also suggests $2\times10^{51}$~erg \citep{laming03}.

\section{The progenitor's stellar wind}
The presence of N-rich ejecta suggests that \casa\ is sweeping up 
CSM from the wind of its progenitor. The swept-up mass is estimated to
be 8~\msun\ \citep{vink96,willingale03}. Such a high mass is
supported by hydrodynamical calculations \citep{laming03,chevalier03},
and by the high ionization parameter (\net) close to the shock front \citep{vink03a},
which suggests an electron density of 30~cm$^{-3}$, which translate into a 
pre-shock H density of 7.5~cm$^{-3}$ or, if the medium is dominated by He, a He density of
3~cm$^{-3}$.
For a uniform wind profile, the density, $\rho$, relates to the mass loss
rate, $\dot{M}$, radius, $r$, and wind velocity, $v_w$, as 
$\rho = \dot{M}/4\pi r^2 v_w$.
Integration to $r=2.55$~pc gives a swept up mass of
$M = 
2.5 (\dot{M}/10^{-3} M_\odot yr^{-1})(v_w/1000\,{\rm km\,s}^{-1})^{-1}$\,\msun.
This means that for a WR wind of typically 2000~\kms, a mass loss rate
of $6.4\times10^{-3}$~\msun/yr is required, which is
in conflict with the observed WR mass loss rates. 
It is, however, more consistent with the slower ($\sim 10-300$~\kms) wind of a 
red super giant (RSG) or LBV star.
This indicates that 
not too long for the explosion (at maximum $10^4$ yr, the expansion time scale
of the optical knots) the progenitor was still in an RSG/LBV phase, placing
strong constraints on the main sequence mass of \casa's progenitor
\citep{garcia96}.

\section{Possibility of enhanced positron escape}
One of the important goals of the \integral\ mission is the galactic
\ep\ annihilation radiation \citep[see][for some first results]{jean03}.
A dominant source for this emission is probably positrons that have escaped
supernovae \citep[e.g.][]{milne01},
where they are produced by the beta decay of  \scff\ and \cofs.
The escape fraction is of importance for the bolometric light curve
of supernovae \citep{ruiz98}.

\casa\ will be observed by \integral, and apart from \tiff, it is interesting
to look for \ep\ annihilation radiation.
The typical annihilation time scale is $10^5$~yr.
The estimated escape fraction for positrons 
produced by \cofs\ decay is $\sim$1\% \citep{chan93}.
Given the \tiff\ yield, this would mean a current annihilation line 
flux of $\sim10^{-6}$~\fluxunit, below the \integral\ sensitivity limits.

However, the conditions that have led to a high \tiff\ yield may also
have favored a high positron escape fraction. Annihilation during the
explosion involves the slowing down of the positrons, mainly through ionization
losses \citep{chan93}. Alpha-rich freeze occurs if the plasma is expanding
rapidly,
but the rapid drop in density also decreases the positron loss and
annihilation rate. Moreover, the ionization losses in the presence of 
He-rich plasma are decreased, due to the small He ionization cross-sections.
Preliminary calculations show that these effects may increase the escape
fraction to 10\%, if most of the $\sim0.07$~\msun\ of $^{56}$Ni 
had initial velocities in excess of 5000~\kms.
This would increase the expected line flux to $\sim10^{-5}$~\fluxunit. 
Unfortunately, the circumstances
that lead to a large escape fraction, will result
in a population of energetic positrons.
The annihilation line may therefore be too much affected by line broadening
to detect with \integral.
Nevertheless, this mechanism may be important, as it potentially
contributes to the galactic positron production \citep{milne02}.

\section{Conclusion}
Observational evidence suggests that \casa\ was a Type Ib SN.
The estimated O mass
of 1-3~\msun\ corresponds to a main sequence mass of 18-25\msun\ (WW95). 
An additional mass constraint is that the progenitor was probably a 
WR star, which puts a lower limit on the main sequence mass of
22\msun\ \citep{massey00}. One should be careful with using the O mass, 
as WW95 also predict a higher than observed Ne and Mg yield. However,
a relatively low mass WR star is supported by a large amount of swept up
mass and the high density behind the blastwave, which suggests that
the progenitor was only briefly in the WR phase.

The mass of the progenitor at the time of 
explosion is difficult to reconcile with
current knowledge of WR stars. X-ray studies suggest an ejecta mass of
less than 4\msun, adding to that the mass of the compact object,
for which the \chandra\ point source is an excellent candidate 
\citep{tananbaum99}, 
suggests a progenitor mass of $\sim$6\msun. 
The fact that we see nucleosynthesis
products from near the core, indicates that most of the ejecta mass must have 
been shocked. This is at odds with current evolutionary models for massive stars 
with rotation,
which suggest that WR star end their lives with about 12\msun\ left
\citep{meynet03}.
Note that Type Ib SN ejecta estimates seem
to agree with that of \casa: 2--4.4\msun\ \citep{hamuy03}.

An alternative scenario, which could explain the low ejecta mass, is binary
mass transfer, but there is no evidence yet 
(e.g. in the form of a runaway star) that the progenitor was part of a 
binary system.
However, the modeling of supernova explosions is an active field of
research, which received a boost from the new found connection between 
Type Ibc supernovae and gamma-ray bursts. For some recent developments and discussions see
for example \citet{heger03} and \citet{kifonidis03}.

\casa's high \tiff\ yield suggests a relatively explosive or an asymmetric
SN event. The X-ray emission
supports both possibilities, with an estimated explosion energy of
$2\times10^{51}$~erg and evidence for a Si-rich jets and fast moving ``plumes''
of Fe-rich plasma. A more compact progenitor can also cause a higher
\tiff\ yield, as a lower ejecta mass results in less
fall back on the stellar remnant.

\vskip 3mm
{\small\noindent\em This work is supported by NASA's 
Chandra Postdoctoral Fellowship Award Nr. PF0-10011
issued by the Chandra X-ray Observatory Center, which is operated by the
SAO under NASA contract NAS8-39073.}


\begin{thebibliography}{48}
\expandafter\ifx\csname natexlab\endcsname\relax\def\natexlab#1{#1}\fi

\bibitem[{{Arnett}(1996)}]{arnett96}
{Arnett}, D. 1996, {Supernovae and Nucleosynthesis} (New Jersey: Princeton
  University Press, 1996)

\bibitem[{{Ashworth}(1980)}]{ashworth80}
{Ashworth}, W.~B. 1980, Jour. Hist. Astron., 11, 1

\bibitem[{{Chan} \& {Lingenfelter}(1993)}]{chan93}
{Chan}, K. \& {Lingenfelter}, R. 1993, \apj, 405, 614

\bibitem[{{Chevalier} \& {Oishi}(2003)}]{chevalier03}
{Chevalier}, R.~A. \& {Oishi}, J. 2003, \apjl, 593, L23

\bibitem[{{Delaney} \& {Rudnick}(2003)}]{delaney03}
{Delaney}, T. \& {Rudnick}, L. 2003, \apj, 589, 818

\bibitem[{{Diehl} \& {Timmes}(1998)}]{diehl98}
{Diehl}, R. \& {Timmes}, F.~X. 1998, \pasp, 110, 637

\bibitem[{{Fesen}(1990)}]{fesen90}
{Fesen}, R.~A. 1990, \aj, 99, 1904

\bibitem[{{Fesen} \& {Becker}(1991)}]{fesen91}
{Fesen}, R.~A. \& {Becker}, R.~H. 1991, \apj, 371, 621

\bibitem[{{Fesen} {et~al.}(2001)}]{fesen01b}
{Fesen}, R.~A. {et~al.} 2001, \aj, 122, 2644

\bibitem[{{Garcia-Segura} {et~al.}(1996){Garcia-Segura}, {Langer}, \& {Mac
  Low}}]{garcia96}
{Garcia-Segura}, G., {Langer}, N., \& {Mac Low}, M.-M. 1996, \aap, 316, 133

\bibitem[{{Gotthelf} {et~al.}(2001)}]{gotthelf01a}
{Gotthelf}, E.~V. {et~al.} 2001, \apjl, 552, L39

\bibitem[{{Hamuy}(2003)}]{hamuy03}
{Hamuy}, M. 2003, \apj, 582, 905

\bibitem[{{Hashimoto} {et~al.}(2001)}]{hashimoto01}
{Hashimoto}, T. {et~al.} 2001, \npa, 686, 591

\bibitem[{{Heger} {et~al.}(2003){Heger}, {Fryer}, {Woosley}, {Langer}, \&
  {Hartmann}}]{heger03}
{Heger}, A., {Fryer}, C.~L., {Woosley}, S.~E., {Langer}, N., \& {Hartmann},
  D.~H. 2003, \apj, 591, 288

\bibitem[{{Hughes} {et~al.}(2000){Hughes}, {Rakowski}, {Burrows}, \&
  {Slane}}]{hughes00a}
{Hughes}, J., {Rakowski}, C., {Burrows}, D., \& {Slane}, P. 2000,
  \apjl, 528, L109

\bibitem[{{Hwang} \& {Laming}(2003)}]{hwang03}
{Hwang}, U. \& {Laming}, J.~M. 2003, ApJ in press, astro-ph/0306120


\bibitem[{{Iyudin} {et~al.}(1994)}]{iyudin94}
{Iyudin}, A.~F. {et~al.} 1994, \aap, 284, L1

\bibitem[{{Jean} {et~al.}(2003)}]{jean03}
{Jean}, P. {et~al.} 2003, \aap, 407, L55

\bibitem[{{Kamper} \& {van den Bergh}(1976)}]{kamper76}
{Kamper}, K. \& {van den Bergh}, S. 1976, \apjs, 32, 351

\bibitem[{{Kifonidis} {et~al.}(2003){Kifonidis}, {Plewa}, {Janka}, \& {M{\"
  u}ller}}]{kifonidis03}
{Kifonidis}, K., {Plewa}, T., {Janka}, H.-T., \& {M{\" u}ller}, E. 2003, \aap,
  408, 621

\bibitem[{{Laming}(2001)}]{laming01c}
{Laming}, J.~M. 2001, in AIP Conf. Proc. 598: Joint SOHO/ACE workshop "Solar
  and Galactic Composition", 411

\bibitem[{{Laming} \& {Hwang}(2003)}]{laming03}
{Laming}, J.~M. \& {Hwang}, U. 2003, ApJ in press,astro-ph/0306119

\bibitem[{{Lawrence} {et~al.}(1995)}]{lawrence95}
{Lawrence}, S.~S. {et~al.} 1995, \aj, 109, 2635

\bibitem[{{Markert} {et~al.}(1983){Markert}, {Clark}, {Winkler}, \&
  {Canizares}}]{markert83}
{Markert}, T.H., {Clark}, G., {Winkler}, P., \& {Canizares}, C. 1983,
  \apj, 268, 134

\bibitem[{{Massey} {et~al.}(2000){Massey}, {Waterhouse}, \&
  {DeGioia-Eastwood}}]{massey00}
{Massey}, P., {Waterhouse}, E., \& {DeGioia-Eastwood}, K. 2000, \aj, 119, 2214

\bibitem[{{Meynet} \& {Maeder}(2003)}]{meynet03}
{Meynet}, G. \& {Maeder}, A. 2003, \aap, 404, 975

\bibitem[{{Milne} {et~al.}(2002){Milne}, {Kurfess}, {Kinzer}, \&
  {Leising}}]{milne02}
{Milne}, P.~A., {Kurfess}, J., {Kinzer}, R., \& {Leising}, M. 2002,
  \newar, 46, 553

\bibitem[{{Milne} {et~al.}(2001){Milne}, {The}, \& {Leising}}]{milne01}
{Milne}, P.~A., {The}, L.-S., \& {Leising}, M.~D. 2001, \apj, 559, 1019

\bibitem[{{Mochizuki}(2001)}]{motizuki01}
{Mochizuki}, Y. 2001, Nuclear Physics A, 688, 58

\bibitem[{{Mochizuki} {et~al.}(1999){Mochizuki}, {Takahashi}, {Janka},
  {Hillebrandt}, \& {Diehl}}]{motizuki99}
{Mochizuki}, Y., {Takahashi}, K., {Janka}, H.-T., {Hillebrandt}, W., \&
  {Diehl}, R. 1999, \aap, 346, 831

\bibitem[{{Nagataki} {et~al.}(1998){Nagataki}, {Hashimoto}, {Sato}, {Yamada},
  \& {Mochizuki}}]{nagataki98}
{Nagataki}, S., {Hashimoto}, M., {Sato}, K., {Yamada}, S., \& {Mochizuki},
  Y.~S. 1998, \apjl, 492, L45

\bibitem[{{Park} {et~al.}(2002)}]{park02}
{Park}, S. {et~al.} 2002, \apjl, 564, L39

\bibitem[{{Rasmussen} {et~al.}(2001)}]{rasmussen01}
{Rasmussen}, A.~P. {et~al.} 2001, \aap, 365, L231

\bibitem[{{Reed} {et~al.}(1995){Reed}, {Hester}, {Fabian}, \&
  {Winkler}}]{reed95}
{Reed}, J.~E., {Hester}, J.~J., {Fabian}, A.~C., \& {Winkler}, P.~F. 1995,
  \apj, 440, 706

\bibitem[{{Ruiz-Lapuente} \& {Spruit}(1998)}]{ruiz98}
{Ruiz-Lapuente}, P. \& {Spruit}, H.~C. 1998, \apj, 500, 360

\bibitem[{{Ryle} \& {Smith}(1948)}]{ryle48}
{Ryle}, M. \& {Smith}, G. 1948, \nat, 162, 462

\bibitem[{{Stephenson} \& {Green}(2002)}]{stephenson02}
{Stephenson}, F.~R. \& {Green}, D.~A. 2002, {Historical supernovae and their
  remnants} (Oxford: Clarendon Press)

\bibitem[{{Tananbaum}(1999)}]{tananbaum99}
{Tananbaum}, H. 1999, \iaucirc, 7246, 1

\bibitem[{{Thorstensen} {et~al.}(2001){Thorstensen}, {Fesen}, \& {van den
  Bergh}}]{thorstensen01}
{Thorstensen}, J.~R., {Fesen}, R.~A., \& {van den Bergh}, S. 2001, \aj, 122,
  297

\bibitem[{{Vink}(2003)}]{vink03c}
{Vink}, J. 2003, {\adspr\ in press, (astro--ph/0304176)}

\bibitem[{{Vink} {et~al.}(1998){Vink}, {Bloemen}, {Kaastra}, \&
  {Bleeker}}]{vink98a}
{Vink}, J., {Bloemen}, H., {Kaastra}, J.~S., \& {Bleeker}, J.~A.~M. 1998, \aap,
  339, 201

\bibitem[{{Vink} {et~al.}(1996){Vink}, {Kaastra}, \& {Bleeker}}]{vink96}
{Vink}, J., {Kaastra}, J.~S., \& {Bleeker}, J.~A.~M. 1996, \aap, 307, L41

\bibitem[{{Vink} \& {Laming}(2003)}]{vink03a}
{Vink}, J. \& {Laming}, J.~M. 2003, \apj, 584, 758

\bibitem[{{Vink} {et~al.}(2001)}]{vink01a}
{Vink}, J. {et~al.} 2001, \apjl, 560, L79

\bibitem[{{Willingale} {et~al.}(2003){Willingale}, {Bleeker}, {van der Heyden},
  \& {Kaastra}}]{willingale03}
{Willingale}, R., {Bleeker}, J., {van der Heyden}, K., \& {Kaastra},
  J. 2003, \aap, 398, 1021

\bibitem[{{Willingale} {et~al.}(2002){Willingale}, {Bleeker}, {van der Heyden},
  {Kaastra}, \& {Vink}}]{willingale02}
{Willingale}, R., {Bleeker}, J.~A.~M., {van der Heyden}, K.~J., {Kaastra},
  J.~S., \& {Vink}, J. 2002, \aap, 381, 1039

\bibitem[{{Woosley} {et~al.}(1993){Woosley}, {Langer}, \& {Weaver}}]{wlw93}
{Woosley}, S.~E., {Langer}, N., \& {Weaver}, T.~A. 1993, \apj, 411, 823

\bibitem[{{Woosley} \& {Weaver}(1995)}]{ww95}
{Woosley}, S.~E. \& {Weaver}, T.~A. 1995, \apjs, 101, 181

\end{thebibliography}
\end{document}